# A Pearson Effective Potential for Monte-Carlo simulation of quantum confinement effects in various MOSFET architectures

[1]M.-A. Jaud, [1]S. Barraud, [2]P. Dollfus, [3]H. Jaouen

[1]CEA-LETI, MINATEC, 17 rue des Martyrs, 38054 Grenoble cedex 9, France
[2]Institut d'Electronique Fondamentale, CNRS, Univ. Paris Sud (UMR 8622), 91405 Orsay, France
[3]STMicroelectronics, 850 rue Jean Monnet, 38926 Crolles, France
e-mail: marie-anne.jaud@cea.fr

A Pearson Effective Potential model for including quantization effects in the simulation of nanoscale nMOSFETs has been developed. This model, based on a realistic description of the function representing the non zero-size of the electron wave packet, has been used in a Monte-Carlo simulator for bulk, single gate SOI and double-gate SOI devices. In the case of SOI capacitors, the electron density has been computed for a large range of effective field ($10^5$ V.cm$^{-1}$ $\leq$ E$_{eff}$ $\leq$ $10^6$ V.cm$^{-1}$) and for various silicon film thicknesses (5 nm $\leq$ T$_{Si}$ $\leq$ 20 nm). A good agreement with the Schrödinger-Poisson results is obtained both on the total inversion charge and on the electron density profiles. The ability of an Effective Potential approach to accurately reproduce electrostatic quantum confinement effects is clearly demonstrated.

## 1. Introduction

As MOSFETs are downscaled to nanometric dimensions, ultra-thin body devices are required for an optimal electrostatic channel control. In such devices, quantization effects are likely to have a large impact on both electrostatics and carrier transport properties. Consequently, to accurately investigate electron transport in ultimate MOSFET architectures, the usual semi-classical transport models can no longer be applied and new simulation tools accounting for quantum effects in the electron transport description are becoming of great relevance.

In the last few years, some works investigated the possibility to develop quantum models based on a particle description of transport [1-16]. Given the strong analogy between Wigner and Boltzmann formalisms, the Monte-Carlo (MC) method commonly used for semi-classical transport simulation can be extended to the quantum case by considering the Wigner function as an ensemble of pseudo-particles [1,3]. This approach describes well the wave-like nature of particles and has been first applied to the unidimensional (1D) simulation of double-barrier resonant structures. To treat quantization effects in an inversion channel, one may couple self-consistently the 1D Schrödinger equation solved along the confinement direction with the multi subband Boltzmann transport in the source-to-drain direction including 2D scattering rates [4,5]. This mode-space approach properly accounts for quantization effects in ultra-thin double-gate devices but is computationally intensive and may be difficult to extend to other architectures. Recently, some works combining the two previous methods for studying quantum transport in ultra-scaled double-gate MOSFETs have been published [6,7]. Alternatives to the mode-space approach are the quantum corrected potential methods [8-16] which have been demonstrated as an efficient way for including quantization effects in a semi-classical particle Monte-Carlo simulator. Among these techniques, the Gaussian Effective Potential (GEP) formulation [12-16] is of great interest because it is weakly sensitive to the particle noise inherent in MC simulation and it is an alternative to the Schrödinger-Poisson based effective potential [10] that requires to solving the Schrödinger's equation. As already reported in [14-16], the GEP correction can accurately reproduce Schrödinger-Poisson (SP) integral quantities such as the total inversion charge but fails to correctly model the electron density profiles. The discrepancy between GEP and SP density profiles is particularly important close to the SiO$_2$/Si interfaces. It is thus especially critical in ultra-thin double-gate structures where electron wave functions are affected by two such interfaces.

In the present work, we demonstrate the ability of an original Effective Potential formalism to correctly account for electrostatic quantum confinement effects, i.e. to accurately reproduce the SP electron density profiles. Our Pearson Effective Potential (PEP) formulation has been developed and implemented in a Monte-Carlo code (MONACO) [17]. In this formulation, the representation of the electron wave packet is based on the pre-determined dependence of the Schrödinger's wave functions on both the local electrical field and the silicon film thickness. The results are reported for bulk, single gate SOI and double-gate SOI devices. For the first time to the best of our knowledge, an excellent agreement is obtained between the electron density profiles calculated with the SP model and a quantum corrected Monte-Carlo code that does not solve the Schrödinger equation.

This paper is organized as follows. In section 2, we briefly outline the quantum corrected potential approach. Section 3 highlights and investigates the limitations of the usual Gaussian Effective Potential (GEP). This leads us to develop a novel Pearson Effective Potential (PEP) correction, which is described in details in Section 4. At last, a validation of the PEP approach for various MOS architectures is presented in Section 5.

## 2. Quantum corrected potential approach

The quantum corrected potential concept has been first introduced by Madelung and Bohm [18, 19]. Its aim is to reproduce physical effects due to quantization by modifying the electrostatic potential responsible for the carrier movement. The flowchart of the quantum corrected Monte-Carlo algorithm together with an illustration of the potential and of the electron density as a function of the distance from an oxide/silicon interface along the confinement direction (referred to as $x$-axis) are presented in Fig. 1.

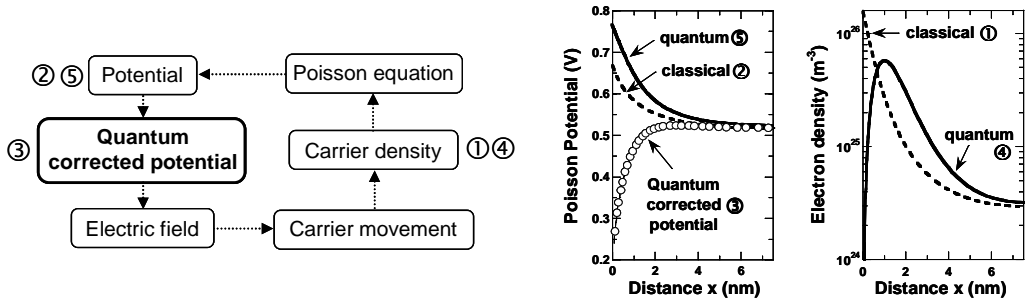

Fig. 1. Principle of the quantum corrected Monte Carlo simulation.

At first, the potential obtained from Poisson's equation solution is used to calculate the quantum corrected potential to be introduced in the Monte-Carlo algorithm for the calculation of carrier trajectories. The resulting quantum corrected potential generates an electric field that tends to repel carriers from oxide/silicon interfaces in accordance with quantization effects. The carrier repulsion at interface is thus naturally included in the standard Monte-Carlo algorithm. As expected, the Poisson's equation solution leads to a "quantum" potential which has a higher curvature than the "classical" potential. Finally, the self-consistency between quantum corrected potential and carrier movement is obtained from an iterative procedure. Within this approach, only the free-flight carrier trajectories are modified by the quantum correction. Scattering mechanisms are assumed to be identical to those of a conventional semi-classical Monte-Carlo approach.

## 3. Gaussian Effective Potential model

### A. Theoretical model

The effective potential formalism has been originally developed by Feynman [20]. It accounts for carrier non-locality by considering the finite size of the carrier wave-packet. As a result, a carrier is not only influenced by the local potential at its position but also by the neighboring potential distribution.

The usual Gaussian Effective Potential (GEP) is defined along the confinement direction by the convolution of the Poisson potential with a Gaussian function representing the electron wave-packet [12,20]:

$$GEP(x) = \frac{1}{\sqrt{2\pi}\,\sigma_x} \int_{-T_{ox}}^{T_{Si}+T_{ox}} V_P(x') \times \exp\left(-\frac{|x-x'|^2}{2\sigma_x^2}\right) dx' \quad (1)$$

where $\sigma_x$ is the standard deviation of the Gaussian function, $T_{Si}$ the silicon film thickness, $T_{ox}$ the oxide thickness and $V_P(x')$ the Poisson potential. As explained in [16], in our code the GEP is calculated using a Fourier transform method. Accordingly, to apply appropriate boundary conditions to the Poisson potential on the oxide areas and to avoid data corruption by convolution in (1), *"Padding regions"* (by reference to signal processing techniques) are to be used on the edge of the device. The parameter $E_B = 3.1$ eV is defined at $SiO_2/Si$ interfaces to represent the oxide barrier height for electrons and satisfies $V_{oxide} = V_P - E_B$.

## B. Results and discussion

As described in [16], we have implemented the GEP correction in the framework of a Monte-Carlo code (MONACO) [17] that uses an analytical conduction-band structure of silicon considering six ellipsoidal nonparabolic $\Delta$ valleys. Double-gate (DG) nMOS capacitors with a channel doping $N_A = 10^{16}$ cm$^{-3}$ and an oxide thickness $T_{ox} = 1$ nm have been simulated. Self-consistent Monte-Carlo simulations corrected by GEP have been performed for a large range of silicon thicknesses (5 nm < $T_{Si}$ < 20 nm) together with a perpendicular effective field $E_{eff}$ varying from $10^5$ V.cm$^{-1}$ to $10^6$ V.cm$^{-1}$. In accordance with [13,15], the standard deviation of the Gaussian function is chosen to be equal to $\sigma_x = 0.5$ nm. Considering the results from SP simulations including the 2-fold and 4-fold valleys with 10 energy levels for each valley as reference, Fig. 2 shows the error on the inversion charge induced by the GEP correction. Fig. 3 compares the electron density resulting from the GEP correction with the one resulting from SP simulation for $T_{Si} = 10$ nm. The GEP formalism is well-known and has been proved to be useful to describe "electrostatic quantum effects" [12-15]. However, errors higher than 10% on the inversion charge are observed at $E_{eff} = 10^5$ V.cm$^{-1}$. At this low effective field, a decrease of the silicon thickness yields a noticeable increase of the inversion charge error (cf. Fig. 2). Moreover, in agreement with [14-15], one can observe in Fig. 3 that the results obtained from Monte-Carlo simulation corrected by the GEP show an overestimated carrier repulsion at $SiO_2/Si$ interfaces. This is due to the fact that the electron wave-packet is systematically represented by a unique Gaussian function, defined by a standard deviation $\sigma_x$ and an average position $R_p$, all along the silicon film thickness. Close to $SiO_2/Si$ interfaces, this description is not realistic with regard to SP results. The inability of the Gaussian function to represent the electron wave-packet has been clearly highlighted in [16] using a methodology based on a design-of-experiments. It has been proved impossible to find out any values of $E_B$ and $\sigma_x$ likely to properly reproduce the SP carrier density profile.

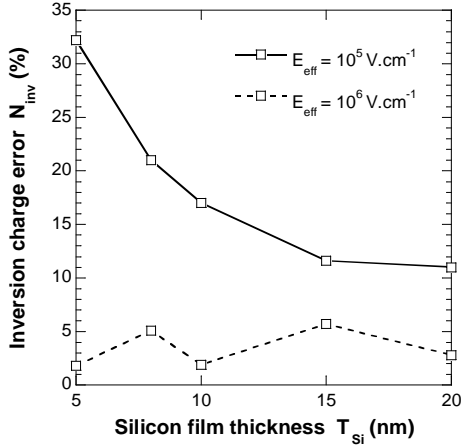 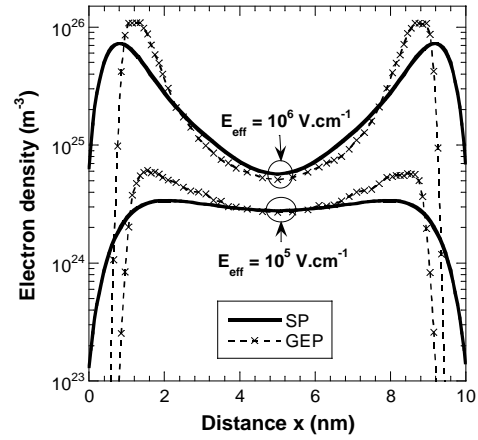

Fig. 2. Inversion charge error in GEP correction (with standard parameters $E_B = 3.1$ eV and $\sigma_x = 0.5$ nm) as a function of the silicon film thickness of double-gate nMOS capacitors for $10^5$ V.cm$^{-1}$ (solid line) and $10^6$ V.cm$^{-1}$ (dotted line) perpendicular effective fields.

Fig. 3. Electron density as a function of the distance in the confinement direction in a double-gate nMOS capacitor with $T_{Si} = 10$ nm using Schrödinger-Poisson (SP - solid lines) and Monte-Carlo corrected by the GEP (GEP – cross dotted lines) models.

## 4. Pearson Effective Potential model

### A. General principle

The previous study based on the GEP correction leads us to propose a new Effective Potential formalism where the electron wave-packet description is improved. The Gaussian function is replaced by a more realistic function based on the shape of the squared modulus of the first level Schrödinger's wave function $|\psi_0|^2$ and carefully calibrated so as to reproduce the electron density profiles resulting from SP simulations considering 10 energy levels. Before calibrating our new function, we first have (i) to choose a well-suited function to reproduce the different possible shapes of $|\psi_0|^2$; (ii) to identify the parameters responsible for the main characteristics of the shape of $|\psi_0|^2$, i.e., to determine the dependences to be given to the new electron wave-packet description. This will lead us to define our novel effective potential formulation.

• *Electron wave-packet's description*

To well describe the various shapes of $|\psi_0|^2$, the new function has to verify the two following conditions: (i) to be a generalization of the Gaussian distribution and (ii) to be possibly asymmetrical. The *Pearson type IV distribution*, often used for the description of doping implantation profiles, fully satisfies these conditions. It is defined by its first four moments which are related to the average position ($R_p$), the standard deviation ($\sigma_p$), the skewness ($\gamma$) and the kurtosis ($\beta$) of the distribution, respectively [21,22] (see Appendix A). Fig. 4 illustrates the influence of each Pearson IV parameter. The skewness and the kurtosis are a measure of the *asymmetry* and *peakedness* of the distribution function, respectively. A positive, respectively negative, value of the skewness results in a maximum of the distribution on the left, respectively on the right, of its average position (cf. Fig. 4b). We can note that a Gaussian function is a particular Pearson IV distribution defined by $\gamma = 0$ and $\beta = 3$.

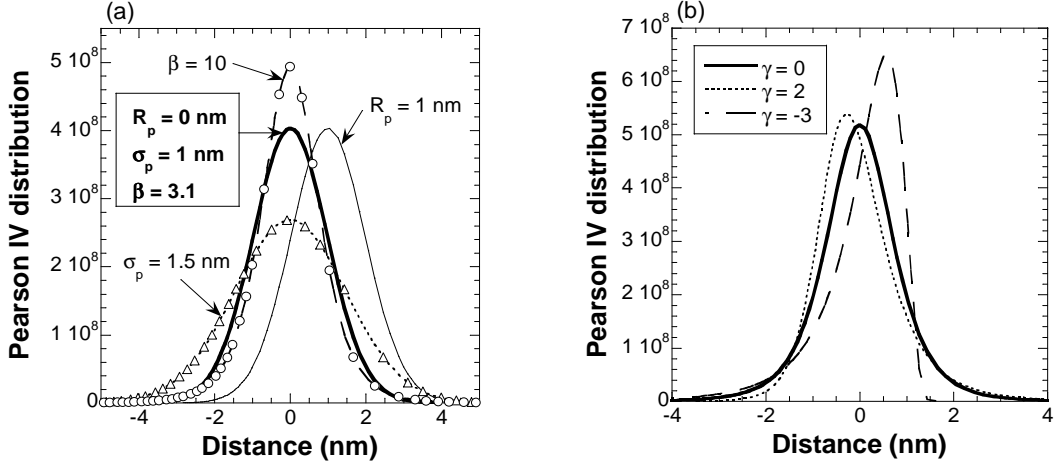

Fig. 4. Pearson IV distributions. (a) $R_p = 0$ nm, $\sigma_p = 1$ nm, $\gamma = 0$, $\beta = 3.1$ (solid heavy line) / $R_p = 1$ nm, $\sigma_p = 1$ nm, $\gamma = 0$, $\beta = 3.1$ (solid line) / $R_p = 0$ nm, $\sigma_p = 1.5$ nm, $\gamma = 0$, $\beta = 3.1$ (open triangles) / $R_p = 0$ nm, $\sigma_p = 1$ nm, $\gamma = 0$, $\beta = 10$ (open circles). (b) $R_p = 0$ nm, $\sigma_p = 1$ nm, $\beta = 30$.

- *Electron wave-packet's dependences*

It is well-known that the shape of $|\psi_0|^2$ is primarily influenced (i) by the potential profile in the confinement direction and (ii) by the silicon film thickness. Therefore, so as to realistically describe the particle wave-packet, Pearson IV parameters should depend (i) on the local electric field $E_x$ in the confinement direction, calculated as the derivative of the potential obtained from Poisson's equation in the confinement direction and (ii) on the silicon film thickness $T_{Si}$. This way, the influence of parameters such as $T_{ox}$, $N_A$ or gate voltage is implicitly taken into account through the $E_x$-dependence.

- *Pearson Effective Potential formulation*

As in the GEP approach, our PEP formulation is based on the convolution of the Poisson potential by a Pearson IV function representing the non zero-size of the electron wave-packet [12,20]. For a DG structure it is defined (1D) as:

$$PEP(x) = \int_{-T_{ox}}^{T_{Si}+T_{ox}} \left[ V_P(x') * \text{Pearson IV } (R_p(E_x, T_{Si}) - x') \right] dx' \quad (2)$$

where $V_P(x')$ is the potential energy, $T_{Si}$ and $T_{ox}$ are the silicon film and oxide thicknesses, and $E_x$ is the local electric field in the confinement direction.

## B. Calibration

To calibrate the four moments of the Pearson IV distribution, the Schrödinger-Poisson equations considering 10 energy levels have been solved self-consistently for double-gate nMOS capacitors with silicon film thickness varying from $5 \text{ nm} \leq T_{Si} \leq 20 \text{ nm}$ and for a large range of effective fields ($10^5$ V.cm$^{-1}$ $\leq E_{eff} \leq 10^6$ V.cm$^{-1}$). Indeed, double-gate capacitors with $T_{Si}$ less than 5 nm are not very realistic for actual technological purposes and the chosen range of effective fields is typical of values used for the effective mobility extraction in the inversion layer of long-channel devices. For each device and effective field, the interfacial electric field, the squared modulus of the first level Schrödinger's wave function $|\psi_0|^2$ and the electron density profile have been extracted. Then, each of the first four theoretical moments of $|\psi_0|^2$ has been calculated as a function of the interfacial electric field and of the silicon film thickness. Thereafter, the terminology "theoretical values" refers to these moment values deduced from SP $|\psi_0|^2$ functions. In the case of a 10 nm film thickness DG capacitor, the theoretical values of the average position with respect to the oxide-silicon interface, the standard deviation and the skewness are plotted in dotted lines as a function of the interfacial electric field on Fig. 5. When decreasing the electric field, the average position is farther away from oxide/silicon interface, the standard deviation is greater and the skewness is smaller, which is in accordance with less pronounced quantum confinement effects.

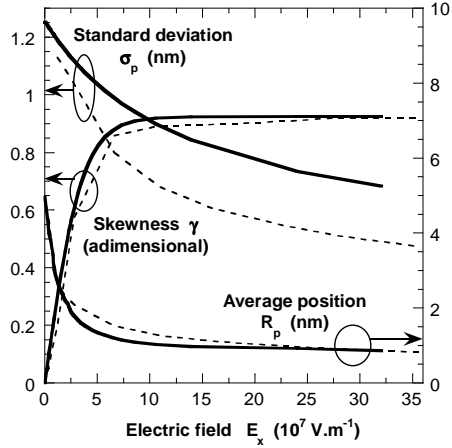

Fig. 5. $R_p$, $\sigma_p$ and $\gamma$ as a function of the electric field $E_x$ in the confinement direction extracted from the squared modulus of the first level Schrödinger's wave function (dotted lines) and defining the Pearson IV distribution of the PEP model (solid lines) for $T_{Si}$ = 10 nm.

The first four moments defining the Pearson IV distributions were calibrated using appropriate functions both to fit theoretical values of $|\psi_0|^2$ as closely as possible and to reproduce SP electron density profiles. The solid lines of Fig. 5 shows the calibration results of average position, standard deviation and skewness obtained for a DG capacitor of 10 nm film thickness. Moreover, for this structure in inversion regime, some Pearson IV distributions associated with various carrier positions in the silicon film as well as the first four moments of the Pearson IV are plotted on Fig. 6 along the confinement direction.

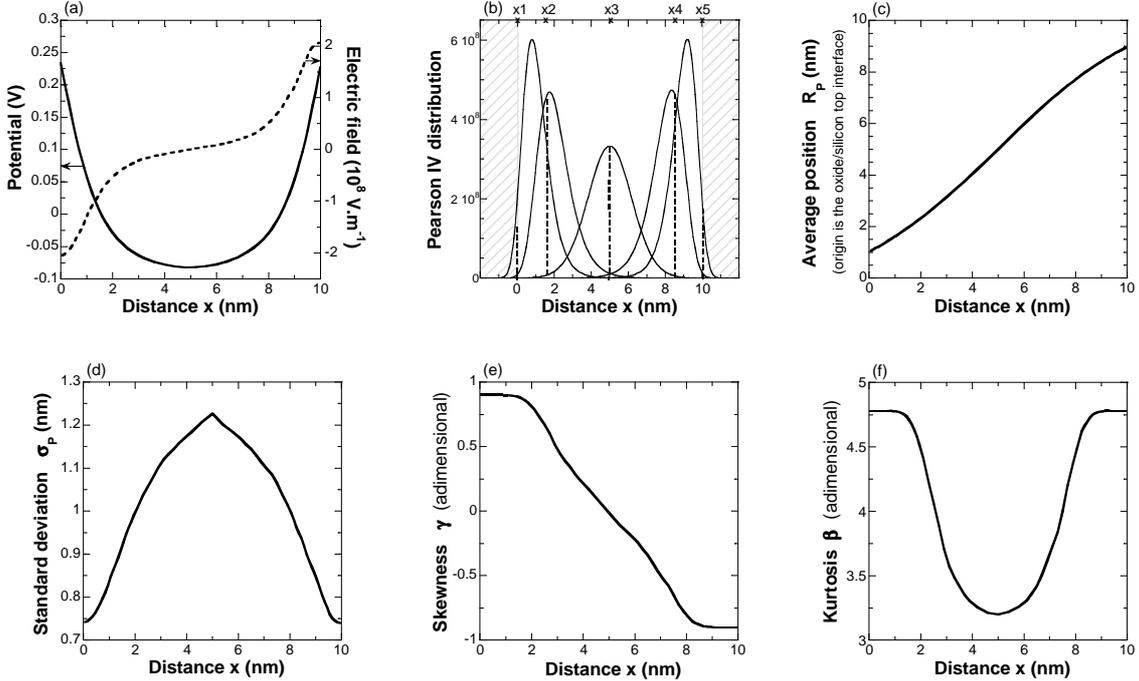

Fig. 6. Poisson potential and local electric field (a), Pearson IV distributions representing the electron wave-packets associated to various electron locations (symbolized by dotted lines) (b), $R_p$ (c), $\sigma_p$ (d) and $\gamma$ (e) and $\beta$ (f) as a function of the distance along the confinement direction for a $T_{Si}$ = 10 nm double-gate nMOS capacitor in inversion regime.

Now we describe in more details the fitting procedure. The expressions of Pearson IV moments as a function of $E_x$ and $T_{Si}$ are given in Appendix B, together with the resulting fitting parameters (Table III).

• For the definition of the average position ($R_p$), the position of the oxide/silicon top interface is taken

as reference. As a function of $E_x$ and $T_{Si}$, $R_p$ is chosen to fit the theoretical values (cf. Fig. 5) while ensuring that (i) in the case of a zero electric field the average position $R_p$ is equal to the particle position and (ii) the $R_p$ evolution along the confinement direction x is continuous and regular, a necessary condition for the numerical stability of the correction algorithm. We can note in Fig. 6c that the average position of the wave-packet of a particle located at oxide/silicon interface is at about 1 nm apart from this interface, which prevents from unrealistic wave-packet penetration in the oxide layer.

• The standard deviation ($\sigma_p$) has been considered as the unique adjustable parameter; i.e. it is not chosen to accurately fit the "theoretical value" but to reproduce the SP electron density profiles. It is explained by the fact that, from SP solution, a weak penetration of the wave-functions in the oxide layer leads to a strong carrier repulsion. In contrast, in Monte-Carlo simulation corrected by an effective potential, a weak penetration of the distribution function assimilated to the particle wave-packet in the oxide layer originates a weak repulsive electric field close to oxide/silicon interfaces, which therefore results in a weak carrier repulsion. That is why the standard deviation of the Pearson IV is not taken identical to the theoretical one but is generally taken slightly higher (cf. Fig. 5). More precisely, $\sigma_p$ is chosen so that the Pearson penetration into the oxide layer induces a repulsive electric field which correctly reproduces electron density profile from SP simulation including several subbands.

• The skewness ($\gamma$) of the Pearson IV distribution has been chosen by fitting the theoretical one (cf. Fig. 5). The sign of the electric field determines the sign of the skewness (cf. Fig. 6e).

• The kurtosis ($\beta$) is arbitrarily calculated as a function of the skewness $\gamma$ so as to be minimal and as close as possible to the Gaussian value [21,22].

Finally, this calibration procedure has allowed us to determine equations defining $R_p$, $\sigma_p$ and $\gamma$ as a function of $E_x$ and $T_{Si}$ as well as $\beta$ as a function of $\gamma$ (see Appendix B). This way, for each carrier position in the confinement direction, the associated Pearson IV distribution is fully defined (cf. Fig. 6b). It can be noted that the Pearson IV representing the wave-packet of a particle located at $SiO_2/Si$ interfaces (x=0=x1 and x=$T_{Si}$=x5) is centred on $R_p \neq x$ and presents a noticeable asymmetry $\gamma \neq 0$. On the other hand, for a particle located at x=$T_{Si}$/2=x3, the Pearson IV looks like a Gaussian function ($\gamma$=0) and is centred on $R_p$=x=$T_{Si}$/2. With our new approach, all along the silicon film thickness and particularly close to the $SiO_2/Si$ interfaces, the particle wave-packet representation is clearly more realistic than a Gaussian distribution. Moreover, since we have calibrated our PEP correction so as to reproduce electron density profiles resulting from SP calculation including 10 energy levels, one can say that our PEP correction integrates the description of valleys and of their associated subbands. However, this technique cannot include the confinement-induced redistribution of electrons among the different valleys as can be done in the Schrödinger-based correction method [10].

### C. PEP calculation flowchart

The generic flowchart of the PEP calculation is presented in Fig. 7. As for the GEP correction, (i) the PEP correction has been implemented in the framework of a Monte-Carlo code (MONACO) [17], (ii) the parameter $E_B$ = 3.1 eV is defined at $SiO_2/Si$ interfaces and satisfies $V_{oxide} = V_P - E_B$. $E_x$ and $T_{Si}$ being known, a set of four parameters ($R_p$, $\sigma_p$, $\gamma$, $\beta$) defining a Pearson IV distribution is calculated at each grid node of the structure as described in the previous section. Let us recall that the solution of Schrödinger's equation is not required for the PEP calculation. The Pearson IV determination only needs the knowledge of calibrated parameters. The Pearson Effective Potential is then calculated at each location "x" as the integral (2) of the product of the Poisson Potential with the associated Pearson IV distribution. Due to the different shapes of the Pearson IV distributions to be considered all along the silicon film thickness, the PEP correction can no longer be performed by a Fourier transform method as in the case of the GEP correction. It is now calculated using a Gaussian quadrature numerical integration method [23].

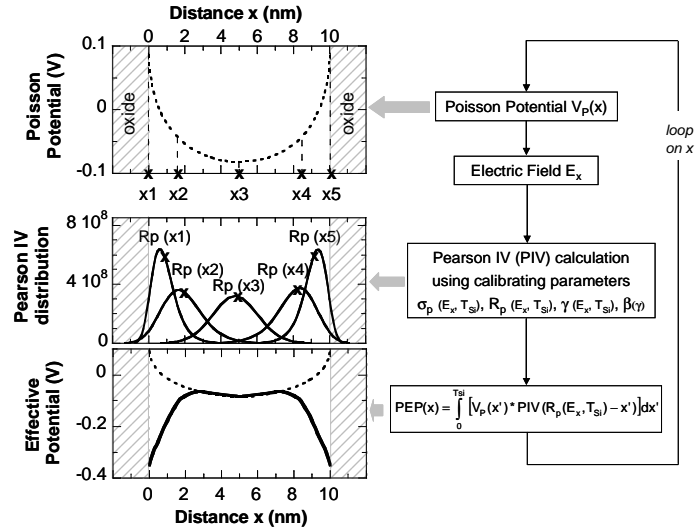

Fig. 7. Flowchart of the Pearson Effective Potential calculation illustrated by results on a double-gate device with $T_{Si} = 10$ nm.

## 5. Pearson Effective Potential electrostatics validation

To validate our original PEP formulation, self-consistent simulations have been performed for several device architectures (double-gate, SOI and bulk). Results of Monte-Carlo simulation corrected by the PEP model are compared with that obtained from SP calculation and from GEP-corrected Monte Carlo simulation (with the value $\sigma_x = 0.5$ nm, as in [13,15]). Because of confinement effects close to both $SiO_2$/Si interfaces, the double-gate nMOS architecture is one of the most critical devices to be tested to assess and demonstrate the ability of our PEP correction to reproduce the SP simulation results. The electron density profiles extracted from double-gate nMOS capacitors with 10 nm silicon film thickness and for a large range of effective fields ($10^5$ V.cm$^{-1}$ ≤ $E_{eff}$ ≤ $10^6$ V.cm$^{-1}$) is shown in Figure 8. While the electron density profiles calculated by the GEP correction are clearly unrealistic close to the Si/SiO$_2$ interfaces due to an unsuitable description of the particle wave-packet, those obtained by the PEP correction agree very well with SP results. Fig. 9 compares the Poisson potential resulting from the PEP correction (open circles) with that resulting from SP simulation (solid line). An excellent agreement is obtained between both approaches. The Poisson potential resulting from semi-classical Monte-Carlo simulation (dotted line) and the Pearson Effective Potential which is actually responsible for the carrier movement (open squares) are also plotted in Fig. 9. As expected the "quantum" Poisson potential exhibits a higher curvature than the "classical" one. Same results have been shown for double-gate nMOS capacitors with an oxide thickness $T_{ox}$ varying from 0.5 nm to 2 nm and a silicon film thickness $T_{Si}$ ranging from 5 nm to 20 nm without any change in the Pearson IV parameters [24].

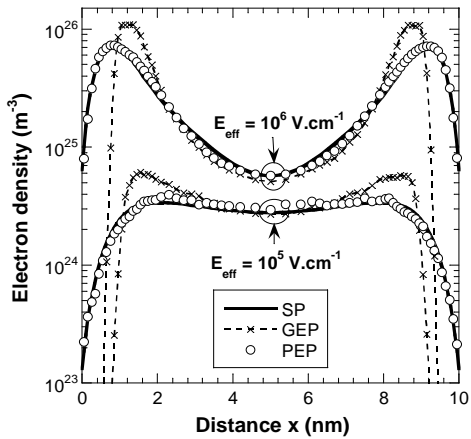 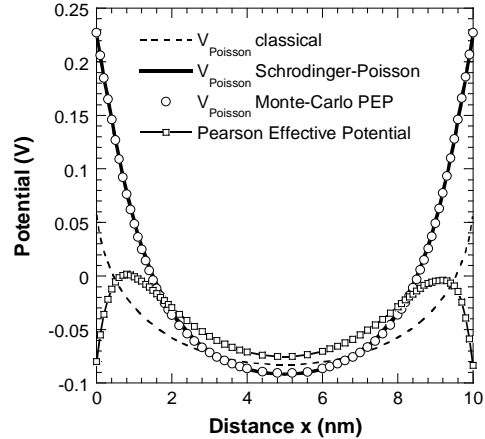

Fig. 8. Electron density as a function of the distance in the confinement direction in a double-gate nMOS capacitor with $T_{Si} = 10$ nm, $T_{ox} = 1$ nm, $N_A = 10^{16}$ cm$^{-3}$ and using SP (solid lines), GEP (cross dotted lines) and PEP (open circles) models.

Fig. 9. Self-consistent Poisson Potential resulting from semi-classical (dotted line), SP (solid heavy line) and Monte-Carlo with PEP correction (open circles) simulations and effective potential (PEP – open squares) as a function of the distance along the confinement direction extracted from a double-gate nMOS capacitor ($T_{ox} = 1$nm, $T_{Si} = 10$nm, $N_A = 10^{16}$cm$^{-3}$).

Results obtained for a 5 nm silicon oxide thickness Silicon On Insulator (SOI) capacitor and bulk nMOS capacitor with a channel doping $N_A = 10^{18}$ cm$^{-3}$ and an oxide thickness $T_{ox} = 1$ nm are presented in Fig. 10 and 11, respectively. The simulations have been performed using the same calibrated parameters as for the DG structure. The electron density resulting from the PEP correction still properly reproduces SP results.

Finally, the ability of the GEP and PEP quantum corrections to conserve the total inversion charge $N_{inv}$ for double-gate (DG), SOI and bulk devices is gathered in Table I. The results of SP simulations are taken as reference. At high effective field, the total inversion charge $N_{inv}$ is accurately reproduced by both approaches. In contrast, at low effective field, the PEP correction generates an error of more than 10% lower than that induced by the GEP. Thus, besides reproducing accurately the SP electron density profiles, the PEP correction also leads to inversion charge errors at the worst equal to the GEP ones or even considerably reduced.

All these results highlight that the PEP correction is well-suited for ultimate bulk, SOI or double-gate nMOS devices with various $T_{Si}$, $T_{ox}$, $N_A$ and gate bias without any additional calibration. This "universality" mainly results from a judicious calibration of Pearson IV parameters as a function of the local electric field in the confinement direction and of the silicon film thickness.

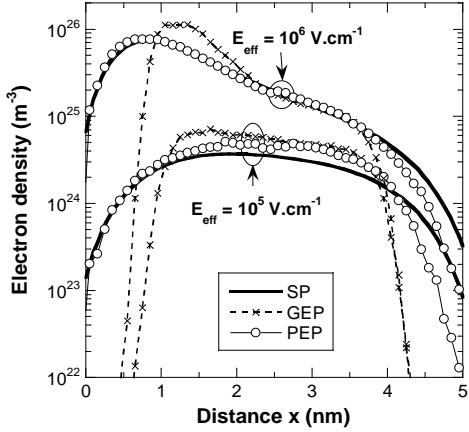 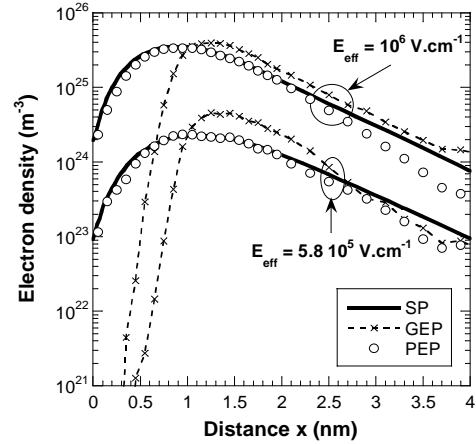

Fig. 10. Electron density as a function of the distance in the confinement direction in a SOI nMOS capacitor with $T_{Si} = 5$ nm, $T_{ox} = 1$ nm, $N_A = 10^{16}$ cm$^{-3}$ and using SP (solid lines), GEP (cross dotted lines) and PEP (open circles) models.

Fig. 11. Electron density as a function of the distance in the confinement direction in a bulk nMOS capacitor with $T_{ox} = 1$ nm, $N_A = 10^{18}$ cm$^{-3}$ and using SP (solid lines), GEP (cross dotted lines) and PEP (open circles) models.

| nMOS capacitor | | | Monte-Carlo GEP | | Monte-Carlo PEP | |
|---|---|---|---|---|---|---|
| Device | $T_{Si}$ (nm) | $T_{ox}$ (nm) | Low $E_{eff}$ | High $E_{eff}$ | Low $E_{eff}$ | High $E_{eff}$ |
| DG | 20 | 1 | 11.0 | 2.8 | 0.7 | 0.4 |
| DG | 15 | 1 | 11.6 | 5.7 | 0.2 | 2.7 |
| DG | 10 | 1 | 17.1 | 1.9 | 4.8 | 1.6 |
| DG | 8 | 1 | 21.0 | 5.1 | 6.8 | 0.8 |
| DG | 5 | 1 | 32.2 | 1.8 | 21.6 | 3.9 |
| DG | 10 | 2 | 13.3 | 1.1 | 2.3 | 1.1 |
| DG | 10 | 0.5 | 23.2 | 4.1 | 5.6 | 2.7 |
| SOI | 10 | 1 | 24 | 3 | 1.6 | 0.7 |
| SOI | 5 | 1 | 35 | 2.3 | 23 | 2.3 |
| Bulk | | 1 | 23.5 | 13.1 | 7.87 | 9.7 |

Table I. Inversion charge error (in percentage) for various nMOS capacitors. Low $E_{eff}$ corresponds to $10^5$ V.cm$^{-1}$ for double-gate (DG) and SOI devices and to $5.8 \times 10^5$ V.cm$^{-1}$ for bulk devices. High $E_{eff}$ corresponds to $10^6$ V.cm$^{-1}$.

# 6. Conclusion

In this work, a new effective potential scheme including properly quantization effects has been developed and implemented into a semi-classical Monte-Carlo simulator. It mainly consists of an improvement of the particle wave-packet description: the Gaussian distribution used in the usual GEP correction is replaced by a Pearson IV distribution that can much better fit the square modulus of the ground subband Schrödinger wave function. Thanks to a judicious calibration of Pearson IV parameters as a function of the local electric field in the confinement direction and of the silicon film thickness, we have demonstrated the ability of the PEP correction to accurately predict electrostatic quantum confinement effects in ultimate bulk, SOI or double-gate nMOS devices without any change in Pearson IV parameters which thus appear to have a universal character. Contrary to the GEP approach, excellent agreements are obtained between SP and PEP electron density profiles for a large range of $T_{Si}$ and $E_{eff}$. The average error calculated on the total inversion charge is similar and reasonable with both quantum corrections at high effective field and is considerably reduced at low effective field when the PEP model is used instead of the GEP one.

## Acknowledgment

This work was supported by the Agence Nationale pour la Recherche through project MODERN (ANR-05-NANO-002).

## Appendix A : Pearson IV definition

The Pearson IV distribution is defined as [21,22]:

$$f(x) = K \left[b_0 + b_1(x-R_P) + b_2(x-R_P)^2\right]^{\frac{1}{2b_2}} \exp\left[-\frac{\frac{b_1}{b_2} + 2b_1}{\sqrt{4b_0 b_2 - b_1^2}} \text{a tan}\left(\frac{2b_2(x-R_P) + b_1}{\sqrt{4b_0 b_2 - b_1^2}}\right)\right] \quad (3)$$

with $b_0$, $b_1$ and $b_2$ given by:

$$b_0 = -\frac{\sigma_P^2 \left(4\beta - 3\gamma^2\right)}{10\beta - 12\gamma^2 - 18} \quad (4)$$

$$b_1 = -\frac{\gamma \, \sigma_P \, (\beta + 3)}{10\beta - 12\gamma^2 - 18} \quad (5)$$

$$b_2 = -\frac{2\beta - 3\gamma^2 - 6}{10\beta - 12\gamma^2 - 18} \quad (6)$$

and K is a constant to ensure that the Pearson IV is normalized.

The skewness γ and the kurtosis β obey the following conditions:

$$0 < \gamma^2 < 32 \qquad \beta > \frac{39\gamma^2 + 48 + 6\left(\gamma^2 + 4\right)^{\frac{3}{2}}}{32 - \gamma^2} \quad (7)$$

We recall that the average position $R_p$, the standard deviation $\sigma_p$, the skewness γ and the kurtosis β are defined as a function of the first four moments of the distribution function as following:

$$R_P = \mu_1 \qquad \sigma_P = \sqrt{\mu_2} \qquad \gamma = \frac{\mu_3}{\mu_2^{3/2}} \qquad \beta = \frac{\mu_4}{\mu_2^2} \quad (8)$$

## Appendix B : Pearson Effective Potential calibration

In our PEP correction, the wave-packet of a particle located in "x" in the confinement direction and under an electric field $E_x$ is represented by a Pearson IV distribution whose moments have been calibrated as a function of $E_x$ and $T_{Si}$. We present here the expressions of each of the four calibrated Pearson IV moments. Table II gathers all the notations specifying their unit and significance. The parameters' values necessary for Pearson moments calculation are listed in Table III.

• *Average position*

The average position is calculated in two different steps. Firstly, the average position of a particle located at the first interface ($R_{P1}$) and at the second interface ($R_{P2}$) are calculated as a function of $E_x$ and $T_{Si}$ so as to fit the theoretical values:

$$R_P = \frac{T_{Si}}{2} - \frac{1}{\log(10^{R_{Pa}})}\left(\frac{T_{Si}}{2} - R_{P\max}\right) \times \log\left[\frac{10^{R_{Pa}} \times |E_x|}{|E_x|_{\max}} + 1\right] \quad (9)$$

Moreover, for a particle under a zero electric field, the average position of its wave-packet ($R_{P0}$) is equal to its location. Secondly, for each particle location, the average position of its wave-packet $R_P$ is calculated from $R_{P0}$, $R_{P1}$ and $R_{P2}$ while ensuring that $R_P(x)$ is continuous and regular:

$$\text{If } x \leq R_{P0} \quad \text{then } R_P = R_{P0} + (R_{P0} - R_{P1} - x_1) \cdot \frac{\tanh\left(\frac{(x - R_{P0})}{R_{Pdiv}}\right)}{\left|\tanh\left(\frac{(x_1 - R_{P0})}{R_{Pdiv}}\right)\right|}$$

$$\text{else } R_P = R_{P0} - (R_{P0} + R_{P2} - x_2) \cdot \frac{\tanh\left(\frac{(x - R_{P0})}{R_{Pdiv}}\right)}{\left|\tanh\left(\frac{(x_2 - R_{P0})}{R_{Pdiv}}\right)\right|} \quad (10)$$

| Name | Unit | Definition |
|---|---|---|
| $\alpha_1$ | m$^{-1}$ | Constant parameter for $\sigma_P$ calculation $\alpha_1 = 10^9$ m$^{-1}$ |
| $\alpha_2$ | m$^{-1}$ | Constant parameter for $\sigma_P$ calculation $\alpha_2 = 17.10^{11}$ m$^{-1}$ |
| $\beta$ | | Kurtosis (cf. eq. 17) |
| $|E_x|$ | V.m$^{-1}$ | Local electric field in the confinement direction |
| $|E_x|_{max}$ | V.m$^{-1}$ | Constant parameter $|E_x|_{max} = 3.5\ 10^8$ V.m$^{-1}$ |
| $\gamma$ | ad. | Skewness (cf. eq. 16) |
| $\gamma_{max}$ | ad. | Parameter for $\gamma$ calculation (cf. Table III) |
| $R_P$ | m | Average position (cf. eq. 14) |
| $R_{Pa}$ | ad. | Parameter for $R_P$ calculation (cf. Table III) |
| $R_{Pdiv}$ | m | Parameter for $R_P$ calculation (cf. Table III) |
| $R_{Pmax}$ | ad. | Parameter for $R_P$ calculation (cf. Table III) |
| $R_{P0}$ | m | Average position of a carrier under a zero electric field $E_x$ |
| $R_{P1}$ | m | Average position of a carrier located at the 1st interface (cf. eq. 13) |
| $R_{P2}$ | m | Average position of a carrier located at the 2nd interface (cf. eq. 13) |
| $\sigma_P$ | m | Standard deviation (cf. eq. 15) |
| $T_{Si}$ | m | Silicon film thickness |
| $T_{Sis}$ | ad. | Parameter for $\sigma_P$ calculation (cf. Table III) |
| x1 | m | Location of the 1st interface |
| x2 | m | Location of the 2nd interface |

Table II. Unit and significance of all the notations used for the calculation of the Pearson IV calibrated parameters (ad. is for adimensional).

| Name | $T_{Si}$ < 10 nm | $T_{Si} \geq$ 10 nm |
|---|---|---|
| $\gamma_{max}$ | $0.03 \times T_{Si}/10^{-9} + 0.6$ | 0.9 |
| $R_{Pa}$ | 5 | Integer part $[0.7 \times T_{Si}/10^{-9} - 2]$ |
| $R_{Pdiv}$ | 6 10$^{-9}$ | $0.4 \times T_{Si} + 2\ 10^{-9}$ |
| $R_{Pmax}$ | $-0.034 \times T_{Si} + 1.17\ 10^{-9}$ | 0.83 10$^{-9}$ |
| $T_{Sis}$ | $T_{Si}/10^{-9}$ | 10 |

Table III. Values of the parameters as a function of $T_{Si}$ used for Pearson IV calibrated parameters calculation according to the units defined in Table II.

• *Standard deviation*

For a particle under a local electric field in the confinement direction $E_x$, the standard deviation of the Pearson IV representing its wave-packet is calculated as follows:

$$\sigma_P = \frac{1}{\alpha_1} \times \left[\log(T_{Sis}) + \frac{T_{Sis} + 1.5}{50}\right] - \frac{1}{\alpha_2} \times (T_{Sis})^3 \times \log\left[\frac{(-8 \times T_{Sis} + 90) \times |E_x|}{|E_x|_{max}} + 1\right] \quad (11)$$

- *Skewness*

The skewness is calculated as a function of $E_x$ and $T_{Si}$ so as to fit the theoretical values:

$$\gamma = \gamma_{max} \times \tanh\left[\frac{T_{Si}}{10^{-9}} \times \frac{|E_x|}{|E_x|_{max}}\right] \qquad (12)$$

Moreover, the sign of the skewness is then adjust to be in adequacy with the sign of the local electric field in the confinement direction $E_x$.

- *Kurtosis*

In accordance with Pearson IV definition [21,22], the kurtosis is only calculated as a function of the skewness $\gamma$ so as to be minimal and closest to the Gaussian value:

$$\beta = \frac{39\gamma^2 + 48 + 6(\gamma^2 + 4)^{\frac{3}{2}}}{32 - \gamma^2} + \varepsilon \qquad (13)$$

with $\varepsilon > 0$ to prevent from numerical difficulties.